\documentclass[reprint,superscriptaddress,preprintnumbers,amsmath,amssymb,aps,prd,tightenlines,longbibliography]{revtex4-2}

\usepackage{graphicx}
\usepackage{xcolor}
\usepackage[sort&compress]{natbib}
\usepackage{amsmath,amssymb,bm,bbm,slashed,amsfonts}
\usepackage{xr-hyper}
\usepackage[colorlinks=true
,urlcolor=blue
,anchorcolor=blue
,citecolor=blue
,filecolor=blue
,linkcolor=red
,menucolor=blue
,linktocpage=true
,pdfproducer=medialab
,pdfa=true
]{hyperref}
\usepackage{cleveref}
\usepackage{enumerate}
\usepackage{epsfig, subfigure}
\usepackage{setspace}
\usepackage{booktabs, tabularx}
\usepackage{units}
\usepackage{placeins}
\usepackage{multirow}
\usepackage{mathtools}
\usepackage[normalem]{ulem}

\begin{document}

\preprint{APCTP-Pre2026-006}

\title{Kinetic Isocurvature Perturbation} 
 

\author{Kyu Jung Bae}
\affiliation{Department of Physics, Kyungpook National University, Daegu 41566, Korea}

\author{Dhong Yeon Cheong}
\affiliation{Enrico Fermi Institute, Kavli Institute for Cosmological Physics, Leinweber Institute for Theoretical Physics, Department of Physics, The University of Chicago, Chicago, IL 60637, USA}

\author{Jinn-Ouk Gong}
\affiliation{Department of Science Education, Ewha Womans University, Seoul 03760, Korea}
\affiliation{Asia Pacific Center for Theoretical Physics, Pohang 37673, Korea}

\author{Keisuke Harigaya}
\affiliation{Enrico Fermi Institute, Kavli Institute for Cosmological Physics, Leinweber Institute for Theoretical Physics, Department of Physics, The University of Chicago, Chicago, IL 60637, USA}
\affiliation{Kavli Institute for the Physics and Mathematics of the Universe (WPI),
The University of Tokyo Institutes for Advanced Study,
The University of Tokyo, Kashiwa, Chiba 277-8583, Japan}

\author{Chang Sub Shin}
\affiliation{Department of Physics and Institute for Sciences of the Universe, Chungnam National University, Daejeon 34134, Korea}
\affiliation{School of Physics, Korea Institute for Advanced Study, Seoul 02455, Korea}
\affiliation{ Particle Theory and Cosmology Group, Center for Theoretical Physics of the Universe, Institute for Basic Science (IBS), Daejeon 34126, Korea}

\begin{abstract}
We formulate a new class of primordial perturbations called \textit{kinetic isocurvature perturbations}, where the mass density of dark matter is constant relative to the photon number density while the kinetic energy of dark matter fluctuates in space.  
Such perturbations naturally arise in scenarios where a nonrelativistic heavy field decays into relativistic dark matter particles with a spatially modulated rate. 
As dark matter cools and becomes nonrelativistic, these fluctuations in kinetic energy leave large-scale density perturbations essentially unaffected and therefore evade the Cosmic Microwave Background bounds on isocurvature perturbations, yet survive as spatial variations in the free-streaming scale, resulting in patch-by-patch variation of the matter power spectrum.
\end{abstract}

\maketitle

\noindent {\bf Introduction.}
%
%
Understanding the microphysics of dark matter remains one of the key questions of our universe. 
While $\Lambda$CDM successfully addresses our late-time universe on large scales~\cite{Planck:2018vyg}, the early-time dynamics remains to be singled out. 
This includes scenarios in which dark matter can carry a nonnegligible momentum~\cite{Dodelson:1993je, Colombi:1995ze, Shi:1998km, Bode:2000gq, Abazajian:2001nj, Hasenkamp:2012ii,Merle:2013wta, Merle:2015oja, Konig:2016dzg, Amin:2022nlh, Banerjee:2023utz, Liu:2024pjg, Ling:2024qfv, Amin:2025dtd, Liu:2025lts, Harigaya:2025pox, Vogel:2025aut, Amin:2025ayf, Gorghetto:2025uls}. 
The resultant free-streaming properties of dark matter can lead to small-scale features~\cite{Bond:1980ha, 1983ApJ...274..443B, Bode:2000gq, Abazajian:2005xn, Abazajian:2006yn, Viel:2013fqw, Horiuchi:2013noa, Abazajian:2014gza, Murgia:2017lwo, Miller:2019pss, Zelko:2022tgf, Vogel:2022odl, Gilman:2025fhy} that are actively searched through, e.g., Lyman-$\alpha$ forest surveys~\cite{Viel:2005qj,Irsic:2017ixq,Decant:2021mhj, Villasenor:2022aiy}.

Primordial isocurvature perturbations also provide a complementary handle on accessing the physics of dark sector. 
Conventional dark matter isocurvature scenarios usually involve the fluctuation in number density~\cite{Bucher:1999re}, which is constrained through, e.g.,~the Cosmic Microwave Background (CMB) ~\cite{Planck:2018jri}. 
Here we point out that dark matter can instead carry a qualitatively distinct isocurvature component sourced through the initial kinetic energy fluctuations. 
These fluctuations give rise to momentum dispersions that retain patch-dependent fluctuations inherited at production, while the comoving number density evolves adiabatically.

This new class of perturbations, which we call \emph{kinetic isocurvature perturbations}, can satisfy the current CMB and large-scale constraints~\cite{Beutler:2011hx, Ross:2014qpa,BOSS:2016wmc, Planck:2018jri} while leaving observational imprints. 
As dark matter becomes nonrelativistic, the overall magnitude of the energy density perturbations sourced through kinetic isocurvature perturbations redshift away and become cosmologically irrelevant on large scales. 
However, as the initial momentum fluctuations control the free-streaming scale, the perturbations survive as spatial variations of the free-streaming cutoff in the matter power spectrum. 
This leads to distinctive long-wavelength fluctuations of the small-scale power spectrum.

In this Letter we develop a general framework for kinetic isocurvature perturbations. 
We present a concrete scenario 
where dark matter is produced through the decay of a heavy field, with the modulating decay rate producing large momentum fluctuations that naturally generate kinetic isocurvature perturbations. We then discuss the observational correlations in small-scale power spectra. 
We quantify the long-wavelength correlation of the modulation of the position-dependent small-scale matter power spectrum.

\vspace{0.2cm}
\noindent {\bf Kinetic Isocurvature Perturbations.}
%
A generic dark matter component can exhibit two qualitatively distinct isocurvature modes.
Consider a dark matter species with a mass $m_{\rm DM}$, a typical momentum $p_{\rm DM}$, and a number density $n_{\rm DM}$. The energy density of dark matter $\rho_{\rm DM} $ is
\begin{align}
    \rho_{\rm DM} = \sqrt{m_{\rm DM}^2 + p_{\rm DM}^2} n_{\rm DM} \equiv E_{\rm DM}\, n_{\rm DM}. 
\end{align}
Then, the perturbation of $\rho_{\rm DM}$ can be written as 
\begin{align}
    \frac{\delta \rho_{\rm DM}}{\rho_{\rm DM}}
        = \frac{\delta n_{\rm DM}}{n_{\rm DM}}
        + \frac{p_{\rm DM}^2}{E_{\rm DM}^2}
          \frac{\delta p_{\rm DM}}{p_{\rm DM}}.
\end{align}
The first term is the standard isocurvature fluctuations in the number density.  
The second term encodes fluctuations in the momentum (or velocity) distribution. 
Although this contribution is negligible in cold dark matter scenarios, it can be sizable in warm dark matter (WDM) scenarios.

We consider the perturbation modes with $\delta n_{\rm DM}$=0 and $\delta p_{\rm DM} \neq 0$, which we call \emph{kinetic isocurvature perturbations}.
Such fluctuations may be produced when dark matter is produced relativistically from the decay of a heavier particle with a spatial modulation in the decay rate.
In this case the comoving number density evolves adiabatically, while the local momentum distribution retains patch-dependent fluctuations. 
As $p_{\rm DM}(a)$ redshifts, dark matter becomes nonrelativistic. Consequently, the associated kinetic contribution to $\delta\rho_{\rm DM}$ becomes irrelevant when the CMB scale enters the horizon and easily satisfies the CMB bounds~\cite{Planck:2018jri}.

A convenient diagnostic of these kinetic distributions is the free-streaming scale~\cite{Choi:2015yma},
\begin{align}
    \lambda_{\rm FS}
        &\equiv 
        \int_{t_i}^{t_{\rm eq}} dt\,\frac{v(t)}{a(t)}   = \int_{a_i}^{a_{\rm eq}}
          \frac{da}{a^2 H(a)}\,
          \frac{p_{\rm DM}(a)}{E_{\rm DM}(a)}.
\end{align}
Because the \emph{initial} momentum distribution governs the subsequent free-streaming behavior, the patch-dependent modulation of the initial momentum induces spatial fluctuations in $\lambda_{\mathrm{FS}}$, 
\begin{align}
   \delta_{\lambda} \equiv \frac{\delta \lambda_{\rm FS}}{\lambda_{\rm FS}},
\end{align}
leading to a spatially modulated small-scale cutoff in the matter power spectrum.

These isocurvature modes imprint a distinctive \emph{large-scale modulation of small-scale structure}, providing a novel type of large-scale correlation that evades conventional bounds and is regarded as a smoking-gun signal for the kinetic isocurvature perturbations.

\vspace{0.2cm}
\noindent {\bf A Concrete Setup.}
%
We present a concrete realization that produces kinetic isocurvature perturbations. 
An adiabatic number density associated with a modulated kinetic distribution may arise naturally when dark matter is produced from the decay of a heavier particle. 
We consider a massive scalar field $\phi$ which is a subdominant component in the energy budget, decaying into a light dark matter particle $\chi$ with mass $m_{\chi} \ll m_{\phi}$ via 
\begin{align}
    \mathcal{L} \supset 
    \begin{cases}
         A\, \phi \chi^{2} ~~~~(\chi:\mathrm{scalar})
         \\[3pt]
        y\, \phi \bar{\chi}\chi ~~~~(\chi:\mathrm{fermion}) 
    \end{cases},  
\end{align}
where $A$ constitutes a mass dimension and $y$ is dimensionless. 
The corresponding decay rate $\Gamma$ is 
\begin{align}
    \Gamma = 
    \begin{cases}
        \displaystyle \frac{A^2 }{16\pi m_{\phi}}
        \left(1 - \frac{4m_{\chi}^2 }{m_{\phi}^2 } \right)^{1/2}       ~~~~(\chi:\mathrm{scalar})
        \\[12pt]
        \displaystyle \frac{y^2  m_{\phi} }{8\pi }
        \left(1 - \frac{4 m_{\chi}^2 }{m_{\phi}^2 } \right)^{3/2}    ~~~~~~(\chi:\mathrm{fermion}) 
    \end{cases}.
\end{align}
After $\phi$ decays, $\chi$ is initially relativistic and the energy density $\rho_{\chi}$ scales as $a^{-4}$, then turns nonrelativistic once the momentum becomes below $m_{\chi}$. 
The momentum of dark matter leads to a free-streaming behavior over a length scale $\lambda_{\mathrm{FS}}$.

The modulation in the coupling constants $A$ or $y$ induces a perturbation in the decay rate $\delta \Gamma$, producing perturbations $\delta \rho_\chi$ in the energy density of $\chi$, 
which in turn leads to a perturbation $\delta \lambda_{\rm FS}$ in the free-streaming length. 
This can be achieved when an additional field $\sigma$ controls these parameters. 

Because the number density of $\phi$ is independent of $\sigma$, this setup gives $\delta n_{\mathrm{DM}} / n_{\mathrm{DM}} =0$ while $\delta_{\lambda}\neq 0$. 
This is in contrast with other WDM scenarios.
For example, for thermally produced WDM, 
both $n_{\rm DM}$ and $\lambda_{\mathrm{FS}}$ are adiabatic.
There are WDM models where $\lambda_{\mathrm{FS}}$ can depend on the field value of a light field so that $\delta_\lambda \neq 0$ (e.g.,~\cite{Co:2017mop,Bodas:2025eca}), but $n_{\rm DM}$ also depends on the field value and $\delta n_{\rm DM} \neq 0$, making this subject to dark matter isocurvature bounds.

We will show that $\delta_{\lambda}\lesssim \mathcal{O}(1)$ is allowed without violating the CMB bound on the isocurvature perturbations.
The key aspect is that for $\lambda_{\rm FS}$ satisfying the warmness constraint, even if the kinetic energy of dark matter has a large density contrast such that $\delta_\lambda$ is large, it becomes small by the time the CMB scale enters the horizon.

\begin{figure}[t!]
    \centering
    \includegraphics[width=0.95\linewidth]{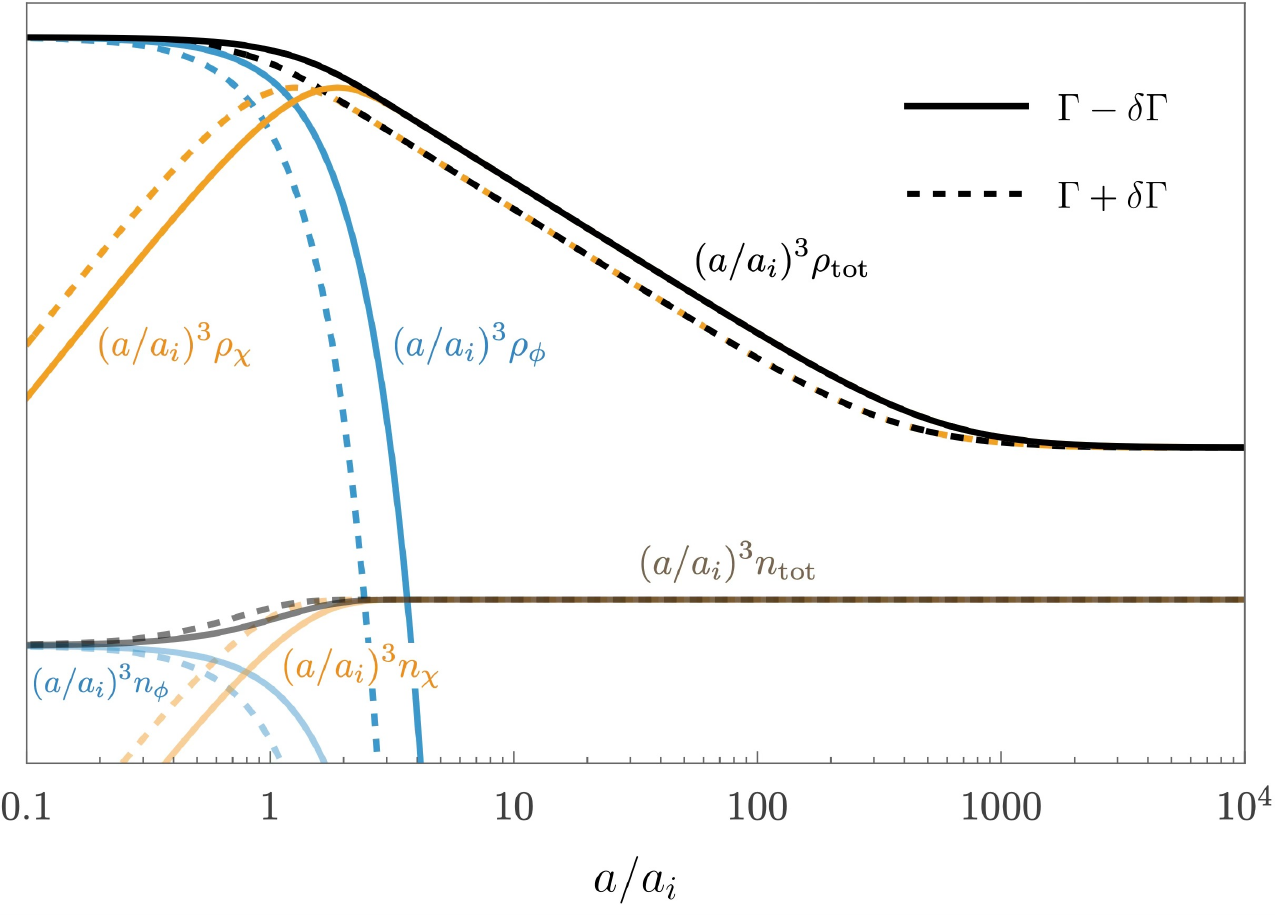}
    \caption{Evolution of different density components in terms of $a/a_{i}$ for $m_{\chi} = 10^{-3} m_{\phi}$ and $\delta \Gamma = 0.2 \Gamma$. The blue lines show $\rho_\phi$ and $n_{\phi}$, the orange ones show  $\rho_\chi$ and $n_{\chi}$, and the black ones depict the total density.}
    \label{fig:densityevol}
\end{figure}

We show how $\delta \rho_{\chi}$ is produced from $\delta \Gamma$. 
Defining the scale factor $a(t_{i}) \equiv a_{i} $ at the time $t_{i} \equiv 1 / \Gamma$, Fig.~\ref{fig:densityevol} represents the density components $(a/a_{i})^{3}\rho$ and $(a/a_{i})^{3} n$ evaluated through the Boltzmann equation provided in the Supplementary Material. 
%
With 
the momentum of dark matter $\chi$ at $a_{i}$ being 
$p_{i} \equiv p(a_{i}) \sim {m_{\phi}}/{2} $, 
the momentum then redshifts as $p(a) = p_{i} \left( {a_{i}}/{a} \right)$.
The modulation of $\Gamma$ leads to that of $p$.
On the other hand, because $\big[a^3 n_{\chi} (a)\big]_{a \gg a_i} = \big[2 a^3 n_{\phi} (a)\big]_{a \ll a_i}$ is independent of the timing of the decay, the contribution of $\delta n_{\mathrm{DM}} / n_{\mathrm{DM}}$ to the isocurvature fluctuation is absent.

We derive an analytic relation between $\delta \rho_\chi$ and $\delta_\lambda$ using the instantaneous-decay approximation.
The density contrast induced by $\delta \Gamma$ is 
\begin{align}
\frac{\delta \rho_{\chi} }{\rho_{\chi} } \simeq -\frac{1}{2}\left[\frac{(m_\phi^2/4)\left(a_i / a\right)^2}{m_\chi^2+(m_\phi^2/4)\left(a_i / a\right)^2}\right] \frac{\delta \Gamma}{\Gamma} . 
\label{eq:densityperturbationgamma}
\end{align}
Assuming that $\chi$ becomes nonrelativistic during the radiation-dominated epoch (RD),
$\lambda_{\mathrm{FS}}$ is given by 
\begin{align}
    \lambda_{\mathrm{FS}} \simeq \frac{p_{i}}{m_{\chi}} \frac{1}{a_{i}H_{i}} \operatorname{tanh}^{-1} \left[ \frac{m_{\chi} }{\sqrt{m_{\chi}^2 + ({a_{i}^2}/{a_{\mathrm{eq}}^2}) p_{i}^2 }}\right],
    \label{eq:approxfreestreaminglength}
\end{align}
from which we obtain $\delta_{\lambda}$ as 
\begin{align}
        \delta_{\lambda} 
    &\simeq -\frac{1}{2} \left(1 + \frac{K(a_{i})}{ \operatorname{tanh}^{-1}\left[ K(a_{i})\right]} \right) \frac{\delta \Gamma}{\Gamma}
    \label{eq:lambdafsperturbationgamma}
\end{align}
with 
\begin{align}
    K(a_{i}) \equiv \frac{m_{\chi}}{\sqrt{m_{\chi}^2 + \left(a_i^2/a_{\mathrm{eq}}^2\right) (m_{\phi}^2/4) }} .
\end{align}
Through Eqs.~\eqref{eq:densityperturbationgamma} and~\eqref{eq:lambdafsperturbationgamma}, we obtain the following relation between 
$\delta\rho_\chi$ and $\delta_\lambda$:
\begin{equation}
\begin{aligned}
    \frac{\delta \rho_{\chi}}{\rho_{\chi}} \simeq &\left[\frac{(m_\phi^2/4)\left(a_i / a\right)^2}{m_\chi^2+(m_\phi^2/4)\left(a_i / a\right)^2}\right] 
    \\ & \times\left(1 + \frac{K(a_{i})}{ \operatorname{tanh}^{-1}\left[ K(a_{i})\right]} \right)^{-1}  \delta_{\lambda}.
    \label{eq:densitypertlambdafspertrelation1}
\end{aligned}
\end{equation}
In our regime of interest, $\lambda_{\mathrm{FS}}$ is well approximated as~\cite{Choi:2012zna} 
\begin{equation}
\begin{aligned}
      \lambda_{\mathrm{FS}} \simeq 1 \, \mathrm{Mpc} \left(\frac{u_{i}^2 t_i }{10^{6} \, \mathrm{sec} }\right)^{1/2} 
      \left[1 - 0.07 \log \left(\frac{u_{i}^2 t_i }{10^{6} \, \mathrm{sec} } \right) \right],
\end{aligned}
\end{equation}
where $u_{i} \equiv p_{i}/m_{\chi}$. 
During RD, the density perturbations when each $k$ mode enters the horizon are 
\begin{equation}
\begin{aligned}
\label{eq:freestreamingpert_temp_approx}
  \frac{\delta \rho_\chi}{\rho_\chi}   &\simeq   \delta_{\lambda}  \left\{1+ 1.56 g_{*}^{-1/2} \left(\frac{1\, \mathrm{Mpc}}{\lambda_{\mathrm{FS}}}\right)^{2} \right.  
  \\ & 
  ~~~~~ \times\left.\left[ 1 + 0.27 \log \left(\frac{1\,\mathrm{Mpc}}{\lambda_{\mathrm{FS}}} \right) \right] \left( \frac{6.7\, \mathrm{Mpc}^{-1}}{k} \right)^{2} \right\}^{-1} 
\end{aligned}
\end{equation}
with $g_{*}$ denoting the number of relativistic degrees of freedom.
Because of the redshift of the kinetic energy,
for $k \lambda_{\rm FS}\ll 1$, the latter term in the curly brackets dominates and suppresses $\delta \rho_{\chi} / \rho_{\chi}$.

\vspace{0.2cm}
\noindent {\bf Existing Constraints.}
%
The upper bounds on 
$\delta_\lambda$
would arise from the associated constraints on $\delta \rho_\chi$ through Eq.~\eqref{eq:freestreamingpert_temp_approx}. 
We consider CMB, baryon acoustic oscillations (BAO), Lyman-$\alpha$, and the $\mu$- and $y$-distortions in the CMB~\cite{Buckley:2025zgh}. 

In order to relate the constraints with the kinetic isocurvature perturbations, we first specify the shape of the power spectrum of $\delta_{\lambda}$, which depends on the origin of the perturbations. 
To be concrete, we take two limiting cases of the spectrum of $\delta_\lambda$: A delta-function-like one and a flat one, 
\begin{align}
    P_{\delta_{\lambda}}^{(\mathrm{delta})} (k ) &= A_{\delta_{\lambda}} \delta(\log k - \log k_{0}), 
    \label{eq:Pisodelta} 
    \\ 
    P_{\delta_{\lambda}}^{(\mathrm{flat})} (k ) & = A_{\delta_{\lambda}} ,
    \label{eq:Pisoflat}
\end{align}
where $P_{\delta_\lambda}$ is defined via $\langle \delta_{\lambda}(\pmb{k})\delta_{\lambda} (\pmb{k}^{\prime }) \rangle =(2\pi)^{3} \delta^{(3)}(\pmb{k}+\pmb{k}^{\prime})  (2\pi^2)P_{\delta_{\lambda}} (k) / k^3 $. 
When each $k$ mode enters the horizon, the power spectrum of the dark matter kinetic isocurvature perturbations based on Eq.~\eqref{eq:freestreamingpert_temp_approx} is 
\begin{align}
    P_{\delta \rho_{\chi}/\rho_{\chi}} (k) \simeq 2 \times 10^{-4} g_{*}\left(\lambda_{\mathrm{FS}} k\right)^{4}  P_{\delta_{\lambda}}(k), 
\end{align}
where we assume $\lambda_{\rm FS} k \ll 1$.

As each mode enters the horizon, the density perturbations $\delta \rho_\chi$ affect the gravitational potential, leading to the photon and matter perturbations measured in the CMB and Lyman-$\alpha$ forests. 
This is similar to the effects of different isocurvature modes on these observational signatures~\cite{Bucher:1999re, Kawasaki:2011rc, Adshead:2020htj}. 
Among the standard isocurvature modes, we reinterpret the bound on the neutrino density isocurvature (NDI) derived in~\cite{Buckley:2025zgh} to constrain $\delta_{\lambda}$.

We match the amplitude of the kinetic isocurvature perturbations $A_{\mathrm{kin}}$ with that of NDI, $A_{\mathrm{NDI}}$, by the identification $\delta \rho_{\chi} / \rho_{\nu} \sim \delta \rho_{\nu} / \rho_{\nu} $.
This gives an overall density ratio scaling that leads to
\begin{align}
    A_{\mathrm{kin}} = A_{\mathrm{NDI}}\left(\frac{\rho_{\nu}}{\rho_{\chi}}\right)^2
\end{align}
when each mode enters the horizon.
This reinterpretation gives a conservative estimate on the constraints for both delta-function-like and flat spectra of the kinetic isocurvature perturbations. 
In fact,~\cite{Buckley:2025zgh} considers two isocurvature profiles: A delta-function spectrum and a broken-power-law spectrum proportional to $(k/k_{0})^{3} $ for $k <k_{0} $ and to $k^{0}$ for $k>k_{0}$, and presents constraints on $A_{\mathrm{NDI}}$ for each pivot scale $k_{0}$. 
For 
$P_{\delta_{\lambda}}^{(\mathrm{delta})}(k)$, the delta-function NDI spectrum constraints can be directly related, with the magnitude of the constraints rescaled with the ratio of $(\rho_{\nu} / \rho_{\chi})^2 $ at the horizon reentry. 
For the flat spectrum case, the $k^2$-scaling originating from $(\rho_{\nu} / \rho_{\chi})^2 $ for $A_{\mathrm{kin}} / A_{\mathrm{NDI}}$ results in a slope shallower than the $k^3$ broken power law at each pivot $k_{0}$. 
Therefore we can conservatively take the most stringently constrained pivot scale for the flat NDI spectrum and regard it as an estimate on the magnitude of the kinetic isocurvature constraints.

\begin{figure}[t!]
    \centering
    \includegraphics[width=0.92\linewidth]{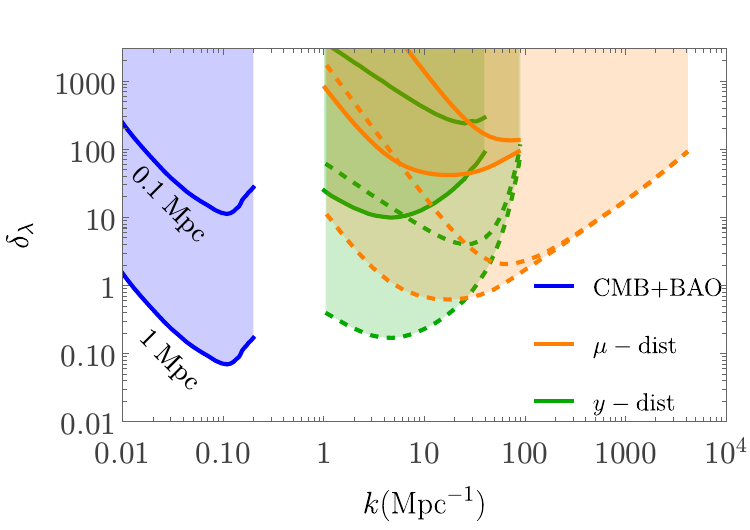}
     \caption{Upper bound on $\delta_{\lambda}$ assuming 
     $P_{\delta_{\lambda}}^{(\mathrm{delta})}(k)$ with the constraints from~\cite{Buckley:2025zgh} for  $\lambda_{\mathrm{FS}}=0.1$ Mpc (top curves) and $1$ Mpc (bottom curves). Solid and dashed lines denote current and prospective bounds, respectively. 
     }
\label{fig:deltalambdaFSdeltafunct}
\end{figure}

We consider the observational constraints for the two representative spectra in Eqs.~\eqref{eq:Pisodelta} and~\eqref{eq:Pisoflat}.
Figure~\ref{fig:deltalambdaFSdeltafunct} shows the current constraints (solid) and future sensitivity (dashed) for 
$P_{\delta_{\lambda}}^{(\mathrm{delta})} (k )$ for $\lambda_{\mathrm{FS}} = 0.1$ and 1 Mpc. 
The low-$k$ region is constrained by CMB+BAO, while the high-$k$ region is constrained by the CMB spectral distortions. The constraints for smaller $\lambda_{\rm FS}$ are weaker because of less kinetic energy of dark matter.
Therefore, for $\lambda_{\mathrm{FS}} \lesssim 0.084~\mathrm{Mpc}$ that satisfy the WDM constraints in the Lyman-$\alpha$ forest~\cite{Harigaya:2025pox}, there are effectively no constraints for perturbative $\delta_\lambda$, allowing for $A_{\delta_{ \lambda}} \lesssim \mathcal{O}(1)$.

\begin{figure}[t!]
    \centering
    \includegraphics[width=0.95\linewidth]{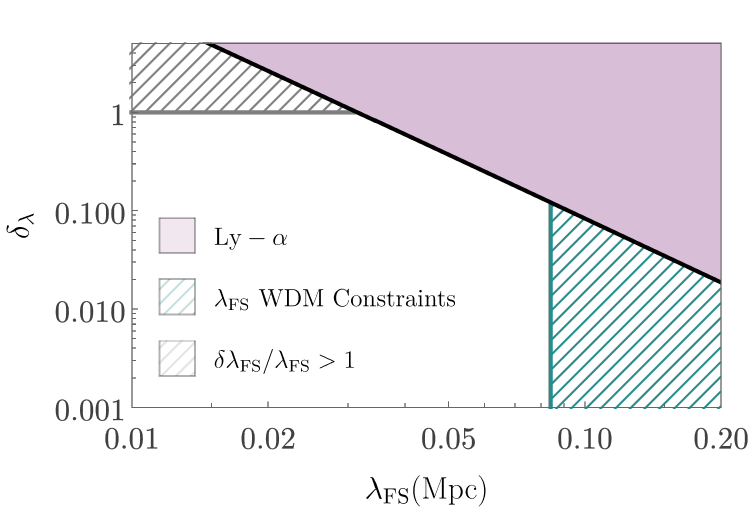}
    \caption{Upper bound on $\delta_{\lambda}$ for given $\lambda_{\mathrm{FS}}$ assuming $P_{\delta_{\lambda}}^{\mathrm{(flat)}}(k)$ with the constraints given in~\cite{Harigaya:2025pox} and~\cite{Buckley:2025zgh}.
    }
    \label{fig:deltalambdaFSupperbound}
\end{figure}

Figure~\ref{fig:deltalambdaFSupperbound} shows the constraints on $\lambda_{\rm FS}$ and $\delta_\lambda$ for 
$P_{\delta_{\lambda}}^{(\mathrm{flat})} (k )$.
The strongest observational constraint arises from the Lyman-$\alpha$ bounds as analyzed in~\cite{Buckley:2025zgh}, with scales of $k\sim 1\,\mathrm{Mpc}^{-1}$ determining the constraints. 
Still, $\delta_\lambda =\mathcal{O}(0.1-1)$ is allowed for $\lambda_{\rm FS}$ that satisfies the warmness bound.
However, we emphasize that the bounds present in Figs.~\ref{fig:deltalambdaFSdeltafunct} and~\ref{fig:deltalambdaFSupperbound} are obtained through a reinterpretation of the NDI bound, and the precise constraints on the kinetic isocurvature perturbations might differ from these values.
In fact, the subhorizon evolution of the kinetic isocurvature perturbations should be different from that of NDI.
We leave a dedicated analysis of the precise evolution for a future work. 
Nonetheless, our conservative estimate on the kinetic isocurvature component tells that current and future observations on isocurvature perturbations are marginally within reach to constrain these kinetic perturbations, and the possibility of a large $\delta_{\lambda} $ generally remains. 
Intriguingly, this magnitude of $\delta_{\lambda} $ would lead to observable large-scale modulation of the small-scale matter power spectrum via the free-streaming effect.

\vspace{0.2cm}
\noindent {\bf Modulating Matter Power Spectrum.}
%
%
A direct way 
$\delta_\lambda$
may manifest itself is through the modulation of the small-scale matter power spectrum $P_{m}(k)$. 
Schematically, this can be observed by measuring the power spectrum within some volume size $\gg \lambda_{\rm FS}^3$ and evaluating the correlation of the fluctuations of the matter power spectra of different patches.

In order to quantify this long-range correlation, we use a position-dependent power spectrum~\cite{Takada:2013wfa, Chiang:2014oga, Chiang:2015pwa} to decouple the long- and short-wavelength behaviors. 
We provide the detailed derivations of the expressions in the Supplementary Material.

Through this, we quantify the small-scale response in the matter power spectrum with respect to the large-scale $\delta_{\lambda} $ over length $\pmb{r}_{L}$ as 
\begin{align}
   \Delta   {P}_{m}(\pmb{k},\lambda_{\mathrm{FS}}(\pmb{r}_L))  &\equiv  
   {P}_{m}\left(\pmb{k}, \lambda_{\mathrm{FS}}\left(\pmb{r}_L\right) \right)- {P}_{m}\left(\pmb{k}, \bar{\lambda}_{\mathrm{FS}} \right) 
   \nonumber\\ 
   & \simeq \lambda_{\mathrm{FS}} \left.\frac{d {P}_{m}}{d \lambda_{\mathrm{FS}}}\right|_{\lambda_{\mathrm{FS}}=\bar{\lambda}_{\mathrm{FS}}}  \delta _\lambda\left(\pmb{r}_L\right)
\end{align}
with $\bar{\lambda}_{\mathrm{FS}}$ denoting the average free-streaming length.
In terms of the momentum $\pmb{K}_L$ conjugate to $\pmb{r}_L$, 
\begin{align}
     P_{PP}(\pmb{k}, \pmb{K}_{L}) =  
     \lambda_{\mathrm{FS}}^2 \left(\frac{d {P}_{m}}{d\lambda_{\mathrm{FS}} } \right)^2   P_{\delta_\lambda}(\pmb{K}_L)
     \label{eq:modulating_matter_power}
\end{align}
with $P_{PP}$, the correlation function of $\Delta{P}_{m}$, defined as $\langle \Delta   {P}_{m} (\pmb{k},\lambda_{\mathrm{FS}};\pmb{K}_{L})   \Delta   {P}_{m}(\pmb{k},\lambda_{\mathrm{FS}};\pmb{K}_{L}^{\prime})\rangle \equiv (2\pi)^{3} \delta^{(3)}(\pmb{K}_{L} + \pmb{K}_{L}^{\prime}) P_{PP}(\pmb{k}, \pmb{K}_{L})$.

\begin{figure}[t!]
    \centering
    \includegraphics[width=0.9\linewidth]{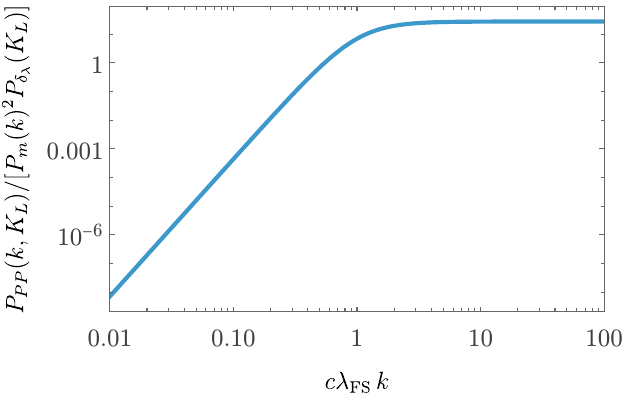}
    \caption{ The ratio $P_{PP}(k,K_{L}) / \big[ P_{m}(k)^2 P_{\delta_{\lambda}}(K_{L}) \big]$ with the transfer function in Eq.~\eqref{eq:transferfunction} with $\beta = 2.4$ and $\gamma = -1.1$. 
    }
    \label{fig:Modulated_power}
\end{figure}

For a representative case, we take the following parameterization for the transfer function~\cite{Murgia:2017lwo}: 
\begin{align}
    T_{\mathrm{FS}}\left(k, \lambda_{\mathrm{FS}}\right)=\left[1+\left(c \lambda_{\mathrm{FS}} k\right)^{\beta}\right]^{\gamma}.
    \label{eq:transferfunction}
\end{align} 
Then the matter power spectrum is expressed as, with $W_s$ being a  window function, $P_{m}(k)=T_{\mathrm{FS}}(k)^2 P_{\mathcal{R}}(k)\left|W_s(k)\right|^2$ with~\cite{Miller:2019pss} $\beta = 2.4$ and $\gamma = -1.1$~\footnote{We emphasize that the precise coefficients for $T_{\mathrm{FS}}$ need a dedicated CLASS analysis. 
For our purposes, it is sufficient to take representative values for WDM produced through particle decay~\cite{Miller:2019pss}. }.
The overall response in the long-range correlation and the matter power spectrum is depicted in Fig.~\ref{fig:Modulated_power}. 
This leads to the overall correlation ratio at  $k = k_{\mathrm{FS}} = 1 / c\lambda_{\mathrm{FS}}$ to be
\begin{align}
    \frac{P_{P P}\left(k_{\mathrm{FS}}, K_L\right)}{P_{m}\left(k_{\mathrm{FS}}\right)^2} \simeq 10 P_{\delta_{\lambda}}\left(K_L\right) .  
\end{align}
%
With $P_{\delta_{\lambda}}^{(\mathrm{flat})}$, 
the large fluctuations allowed for $\delta \lambda $ can induce an $\mathcal{O}(1) $ fluctuation in this matter power spectrum correlation.

This result indicates that distantly separated patches may exhibit the correlation of the fluctuations of the matter power spectrum, which can be revealed by observations that are sensitive to small-scale matter power spectra, such as the galaxy surveys or the observations of the Lyman-$\alpha$ forests.
The measurement of the small-scale power spectrum is usually sensitive to astrophysical systematic uncertainties, but we expect that the spatial modulation helps reduce the uncertainties.

The values of $K_{L}$ must remain hierarchically smaller than the small-scale wavenumber $k$ for the computation of the modulation of $P_{m}(k)$ to be valid. 
Within $K_L$ satisfying this hierarchy, the measurement of $P_{\delta_\lambda}(K_L)$ is affected by different types of observational uncertainties. 
Smaller $K_{L}$ should have lower uncertainties in measuring $P_{m}(k)$ at each patch because of larger volumes, but $P_{\delta_\lambda}(K_L)$ is subject to larger cosmic variance due to the limited number of available independent long-wavelength modes. 
On the other hand, larger $K_{L}$ are less affected by cosmic variance, but the error in determining the power spectrum at each patch will become larger, leading to excessively large errors in measuring $P_{\delta_\lambda}$ for too large $K_L$. 
The determination of the optimal range of $K_L$ in measuring $P_{\delta_\lambda}$ requires careful evaluation of observational uncertainties, but we expect the optimal range to be $\mathcal{O}(0.01-0.1)$ Mpc$^{-1}$, corresponding to scale separations relevant for the CMB and large-scale structure.

\vspace{0.2cm}
\noindent {\bf Conclusions.}
%
In this Letter, we formulated the isocurvature perturbations of the kinetic energy distribution of dark matter. 
In contrast to conventional dark matter isocurvature perturbations, the number density fluctuation is absent. 
We showed that these kinetic perturbations can be generated when dark matter is produced via the decay of a particle whose decay rate is spatially modulated. 
Kinetic perturbations may arise in other settings, which are worth further investigations.

Because these kinetic isocurvature perturbations redshift, the associated CMB-scale isocurvature perturbations are suppressed. 
The strongest current constraints instead arise from small-scale observations, but the fluctuations in the free-streaming length can remain large, potentially reaching \(\mathcal{O}(1)\).
Our results imply that sizable fluctuations may survive in observables that are sensitive to spatial modulations of the small-scale power spectrum. 

Our results open new avenues for probing cosmic perturbations with novel observables. 
A particularly intriguing direction is to identify optimal observables to detect the kinetic isocurvature perturbations.
Candidates for the observable include the Lyman-$\alpha$ forests and galaxy surveys.
We leave a dedicated analysis of observational strategies for future work.

\medskip
\noindent {\it Acknowledgments.}
We thank Kevork Abazajian, Wayne Hu, Austin Joyce, Sarunas Verner, Lian-Tao Wang for useful discussions. 
KJB is supported by the National Research Foundation of Korea grants funded by the Korea government RS-2022-NR070836 and RS-2025-24523746.
DYC is supported by the Enrico Fermi and KICP fellowship from the Enrico Fermi Institute and the Kavli Institute for Cosmological Physics at the University of Chicago, and the Kavli Foundation. 
JG is supported in part by the National Research Foundation of Korea grant funded by the Korea government RS-2024-00336507 and by the Ewha Womans University Research Grants of 2025, 1-2025-0739-001-1 and 1-2025-1260-001-1. 
KH is supported by the Department of Energy grant DE-SC0009924 and the World Premier International Research Center Initiative (WPI), MEXT, Japan (Kavli IPMU). 
CSS is supported by the National Research Foundation of Korea grant funded by the Korea government RS-2025-25442707 and by IBS under the project code IBS-R018-D1.
This work was performed in part at the Aspen Center for Physics, which is supported by National Science Foundation grant PHY-2210452.
Parts of this work were carried out during the academic programs hosted by the Asia Pacific Center for Theoretical Physics, ``Dark Matter as a portal to New Physics 2026'' (APCTP-2026-F01) and ``New perspectives on cosmology 2026'' (APCTP-2026-F02).
JG thanks the Asia Pacific Center for Theoretical Physics for hospitality while this work was under progress.

\bibliographystyle{apsrev4-1}
\bibliography{refs}

\clearpage
\onecolumngrid
\begin{center}
\textbf{\large Kinetic Isocurvature Perturbation} \\ 
\vspace{0.05in}
{ \it \large Supplementary Material}\\ 
\vspace{0.05in}
Kyu Jung Bae, Dhong Yeon Cheong, Jinn-Ouk Gong, Keisuke Harigaya, Chang Sub Shin
{}
{}
\setcounter{equation}{0}
\setcounter{figure}{0}
\setcounter{table}{0}
\setcounter{section}{0}
\setcounter{page}{1}
\end{center}
\makeatletter
\renewcommand{\theequation}{S\arabic{equation}}
\renewcommand{\thefigure}{S\arabic{figure}}
\renewcommand{\thetable}{S\arabic{table}}

\section{Boltzmann equation analysis}
\label{app:Boltzmann}

Here we outline the evolution of the heavy $\phi$ and warm dark matter $\chi$ given in the main text. 
The full evolution of these free-streaming particles can be computed using the Boltzmann equations. 
In principle the metric perturbations will also enter into the evolution equations. 
Given the Newtonian gauge 
\begin{align}
    d s^2=-(1+2 \Phi) d t^2+a^2(1-2 \Psi) \delta_{i j} d x^i d x^j ,
\end{align}
the distribution function $f_{j}$ of a species $j$ follows the Boltzmann equation
\begin{align}
    \frac{\partial f_j}{\partial t}+\frac{1}{E} \frac{\pmb{p}}{a } \cdot \bm{\nabla} { f_j}{}-\frac{p^2}{E} \frac{\partial f_j}{\partial E}\left[H-\frac{\partial \Psi}{\partial t}+\frac{E}{p^2} \frac{1}{a} \,\pmb{p}\cdot \bm{\nabla}{ \Phi}{}\right]=\frac{1}{E}(1+\Phi) \,C\left[f_j\right],
\end{align}
with $C[f_{j}]$ corresponding to the collision terms. 
These perturbations will alter the overall phase space distribution of the WDM $\chi$ particles, affecting the precise form of the transfer function, and will interplay with the modulation features of the matter power spectrum. 
In this work we take the regime where the decay of $\phi$ occurs when the scales of interest are far outside the horizon, for which the separate universe picture holds and the free-streaming perturbation can be taken independently of the metric perturbations. 
Within this limit, the Boltzmann equations for $\phi $ and $\chi $ are expressed as
\begin{align}
    & \frac{\partial f_{\phi} }{\partial t } - H p \frac{\partial f_{\phi} }{\partial p } = \frac{1}{E} C [ f_{\phi} ] ,
    \\
    & \frac{\partial f_{\chi} }{\partial t } - H p \frac{\partial f_{\chi} }{\partial p } = \frac{1}{E}C [ f_{\chi} ] ,
\end{align}
with $C[f_{\phi}]$ and $C[f_{\chi}]$ expressed as
\begin{align}
    C\left[ f_{\phi}  \right] &= - \frac{1}{2} \int \frac{g_\chi d^3{q}_\chi}{(2 \pi)^3 2 E_\chi} \int\frac{g_\chi d^3{q}_\chi^{\prime}}{(2 \pi)^3 2 E_\chi^{\prime}} (2\pi)^{4} \delta^{(4)} (p - q_{\chi} - q_{\chi}^{\prime} ) |\mathcal{M}_{\phi \rightarrow \chi \chi} |^2  
    \Big[ f_{\phi} (\pmb{p},t) - f_\chi  (\pmb{q}_{\chi} ,t)f_{\chi}  (\pmb{q}_{\chi}^{\prime},t) \Big] ,
    \\
    C\left[ f_{\chi}  \right] &=\int \frac{g_\phi d^3{p}_\phi}{(2 \pi)^3 2 E_\phi} \int\frac{g_\chi d^3{q}_\chi^{\prime}}{(2 \pi)^3 2 E_\chi^{\prime}} (2\pi)^{4} \delta^{(4)} (p_{\phi}  - q- q_{\chi}^{\prime} ) |\mathcal{M}_{\phi \rightarrow \chi \chi} |^2  
    \Big[ f_{\phi} (\pmb{p}_{\phi},t) - f_\chi  (\pmb{q},t)f_{\chi}  (\pmb{q}^{\prime},t) \Big] .
\end{align}
This expression imposes CP invariance and the Maxwell-Boltzmann statistics $(1\pm f_{j}) \simeq 1$. 
Ignoring backreaction by neglecting $f_{\chi} f_{\chi} $ terms and taking the $\phi$ distribution function to be 
\begin{align}
    f_{\phi} (\pmb{p},t) = \frac{(2\pi)^{3} }{g_{\phi} }{n}_{\phi} \delta^{(3)} (\pmb{p}),
\end{align}
we find the Boltzmann equations to be 
\begin{align}
\frac{\partial f_{\phi} }{\partial t } - H p \frac{\partial f_{\phi} }{\partial p } &=   - \frac{ A^2 }{16\pi m_{\phi}} \sqrt{1 - \frac{4m_{\chi}^2 }{m_{\phi}^2}}f_{\phi} 
\equiv -\Gamma f_{\phi} , 
\\ 
\frac{\partial f_{\chi} }{\partial t } - H p \frac{\partial f_{\chi} }{\partial p } &= \frac{ \pi  A^2 }{4 m_{\phi} E_\chi^2}\, {n}_{\phi} \,  \delta \bigg( E - \frac{m_{\phi}}{2} \bigg)  ,
\end{align}
where $A^2$ is $|{\cal M}|^2$ averaged over the angular direction. 
In terms of the number density, for a given abundance $Y_{\phi}^{(0)}$, 
\begin{align}
    n_{\phi} & \simeq Y_{\phi}^{(0)} s(T) e^{-\Gamma t}~,
    \\ 
    \rho_{\phi} & \simeq m_{\phi}  Y_{\phi}^{(0)} s(T) e^{-\Gamma t} ~,
\end{align}
where $s(T)$ is the entropy density. During RD, assigning $z\equiv m_{\phi} / T$ and $\xi \equiv p / T$, $f_{\chi} $ takes the form
\begin{align}
    f_{\chi}(\xi, z) = \frac{\pi^{3} A^2 g_{*,s} }{45 \xi^2 m_{\phi} } \frac{Y_{\phi}^{(0)} e^{-\Gamma/ 2H(2\xi) }}{2\xi H(2\xi) }\Theta(z - 2 \xi ).
\end{align}
Integrating over $\xi$ gives the corresponding number density and energy density evolution, which reproduces the evolution in Fig.~\ref{fig:densityevol}.

\begin{figure}[h!]
    \centering
    \includegraphics[width=0.55\linewidth]{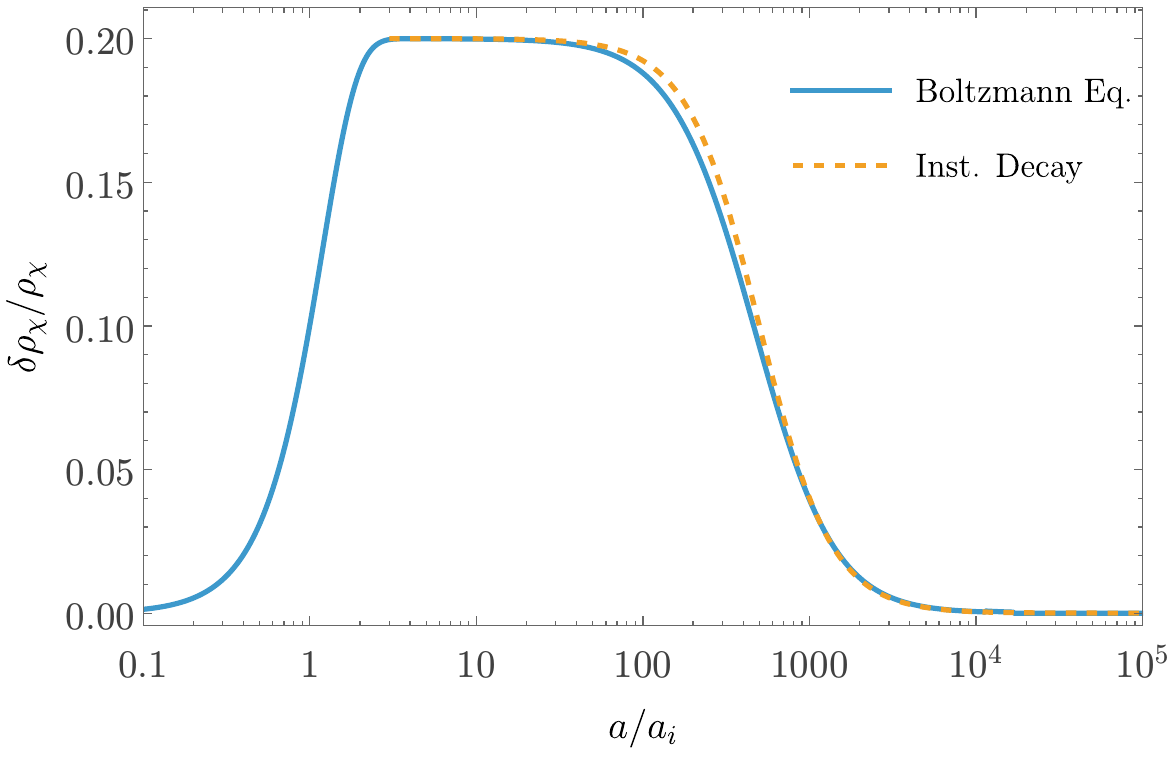}
    \caption{Comparison of $\delta \rho_{\chi} / \rho_{\chi}$ between the result from solving the Boltzmann equation (blue) and that under the instantaneous-decay approximation (orange) for $m_{\chi} = 10^{-3} m_{\phi}$ and $\delta \Gamma = 0.2 \Gamma$.  } 
    \label{fig:density_perturbation_comparison}
\end{figure}

In the main text, we derived the relation between $\delta \rho_\chi/\rho_\chi$ and $\delta \Gamma/\Gamma$ using the instantaneous-decay approximation at $\Gamma = 3H$. 
Figure~\ref{fig:density_perturbation_comparison} shows the comparison of the approximation with the exact result, justifying the validity of the instantaneous-decay approximation. 

\section{Derivation of the Modulating Matter Power Spectrum}
\begin{figure}[h!]
    \centering
    \includegraphics[width=0.7\linewidth]{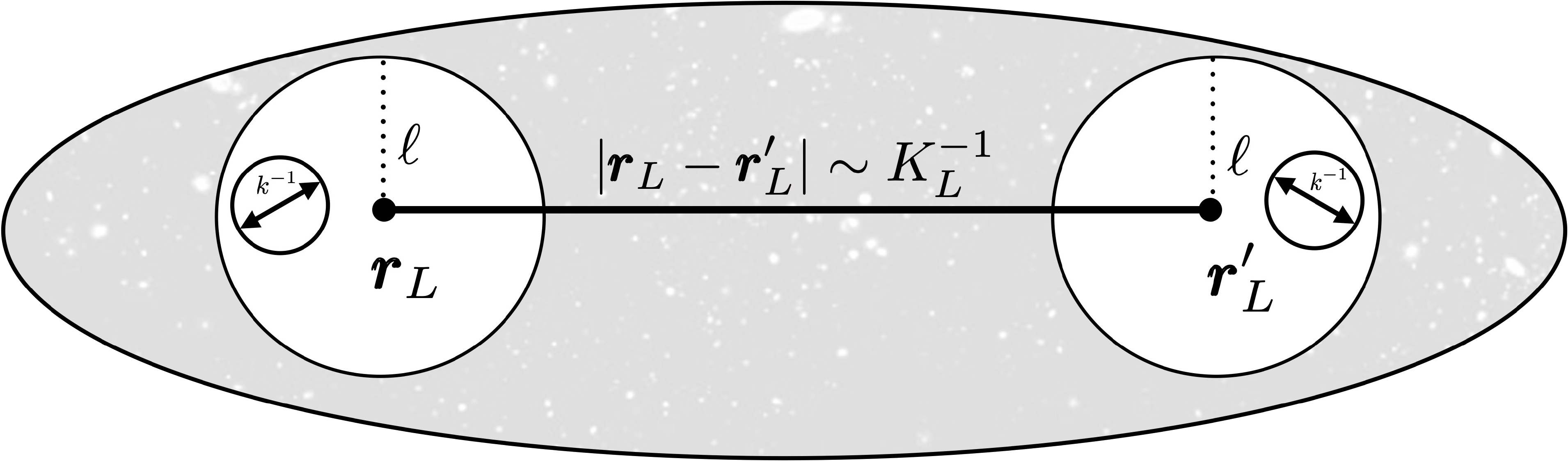}
    \caption{Schematic presentation of the position-dependent power spectrum with corresponding length scales $K_L^{-1} \gg \ell \gg k^{-1}$. } 
    \label{fig:schematic}
\end{figure}
Here we provide explicit derivations of the effects of the kinetic isocurvature perturbations on the long-range correlation of the modulating matter power spectrum in the main text.
A concrete framework is to use the position-dependent power spectrum~\cite{Takada:2013wfa, Chiang:2014oga, Chiang:2015pwa} to formulate this long-range correlation for small scales. 
Figure~\ref{fig:schematic} represents the qualitative picture of the three length scales present in the setup, $K_L^{-1} \gg \ell \gg k^{-1}$.
One can consider the fluctuations within a subvolume $V_{s}$ with length $\ell$ centered at $\pmb{r}_{L}$, and define the local power spectrum $P_\delta (\pmb{k}, \pmb{r}_L)$. 
The local power spectra have spatial fluctuations with long-range correlation over length scales $K_L^{-1} \sim |\pmb{r}_{L}-\pmb{r}_{L}'|$ due to the kinetic isocurvature perturbations.

We now quantitatively show the correlations within this setup by taking a generic perturbation variable denoted as $\delta(\pmb{r})$, which may be taken to be the matter density contrast. 
We identify a subvolume $V_{s}$ to be centered at some position $\pmb{r}_{L}$ and define the position-dependent power spectrum within this local patch as
\begin{align}
    {P}_{\delta}\left(\pmb{k}, \pmb{r}_L\right) &\equiv \frac{1}{V_s}\left|\delta\left(\pmb{k}, \pmb{r}_L\right)\right|^2, 
    \\
        \delta\left(\pmb{k}, \pmb{r}_L\right) &\equiv  \int d^3 r \,\delta(\pmb{r}) W_s\left(\pmb{r}-\pmb{r}_L\right) e^{-i  \pmb{k} \cdot \pmb{r} }
        = \int \frac{d^3 q}{(2 \pi)^3} \delta^{(3)}(\pmb{k}-\pmb{q}) W_s(\pmb{q}) e^{-i \pmb{q} \cdot \pmb{r}_{L}} ,
\end{align}
with $\delta\left(\pmb{k}, \pmb{r}_L\right)$ being the local Fourier transform of the perturbation variable $\delta$ and $W_{s}$ being a window function. 
As long as the scale associated with the small-scale perturbations of interest is smaller than the length scales corresponding to $V_{s}$, the physical features are independent from the exact form of the window function. 

We derive the two-point correlation function of the modulation of the matter power spectrum $P_{m}(\pmb{k}, \pmb{r}_{L})$ from the modulating $\lambda_{\mathrm{FS}}$. 
We express the modulation of $\lambda_{\mathrm{FS}} (\pmb{r}_{L})$ with respect to the globally averaged $\bar{\lambda}_{\mathrm{FS}}$,
\begin{align}
    \lambda_{\mathrm{FS}} (\pmb{x}) = \bar{\lambda}_{\mathrm{FS}} + \delta \lambda_{\mathrm{FS}}(\pmb{x}). 
\end{align}
The survey volume-averaged free-streaming scale then becomes
\begin{align}
    \delta \lambda_{\mathrm{FS}}(\pmb{r}_{L}) = \frac{1}{V_{s}}\int d^{3} r \, \delta\lambda_{\mathrm{FS}}(\pmb{r}) \,W_{s}(\pmb{r} - \pmb{r}_{L}). 
\end{align}

Within the separate universe approach~\cite{Sasaki:1998ug,  Wands:2000dp, Lyth:2004gb, Sirko:2005uz, Baldauf:2011bh, Takada:2013wfa, Li:2014sga, Li:2014jra, Wagner:2014aka}, $\lambda_{\mathrm{FS}}$ will be constant within a subvolume patch. 
We expand the small-scale matter power spectrum in terms of $\delta \lambda_{\mathrm{FS}}$ as
\begin{align}
    P_{\delta }(\pmb{k},\lambda_{\mathrm{FS}}(\pmb{r}_L);\pmb{r}_{L}) 
    = 
    P_{\delta }(\pmb{k},\bar{\lambda}_{\mathrm{FS}};\pmb{r}_{L}) 
    + \left.\frac{d P_{\delta}}{d\lambda_{\mathrm{FS}} } \delta \lambda_{\mathrm{FS}}(\pmb{r}_{L})\right|_{\lambda_{\mathrm{FS}} = \bar{\lambda}_{\mathrm{FS}}} + \cdots. 
\end{align}
Keeping fluctuations up to linear order in $\delta \lambda_{\mathrm{FS}}$, the fluctuation in $P_{\delta }(\pmb{k},\lambda_{\mathrm{FS}}(\pmb{r}_L);\pmb{r}_{L})$ becomes
\begin{align}
    \Delta   P_{\delta }(\pmb{k},\lambda_{\mathrm{FS}}(\pmb{r}_L);\pmb{r}_{L}) 
    = 
    \left.\frac{d P_{\delta}}{d\lambda_{\mathrm{FS}} } \delta \lambda_{\mathrm{FS}}(\pmb{r}_{L})\right|_{\lambda_{\mathrm{FS}} = \bar{\lambda}_{\mathrm{FS}}}.
\end{align}
The correlation function of $\Delta   P_{\delta }$ is
\begin{align}
    \langle   \Delta   P_{\delta}(\pmb{k},\lambda_{\mathrm{FS}}(\pmb{r}_L);\pmb{r}_{L})   
    \Delta   P_{\delta }(\pmb{k},\lambda_{\mathrm{FS}}(\pmb{r}_L^{\prime});\pmb{r}_{L}^{\prime})\rangle 
    = 
    \left(\frac{d P_{\delta}}{d\lambda_{\mathrm{FS}} } \right)^2 
    \langle \delta \lambda_{\mathrm{FS}}(\pmb{r}_{L})\delta \lambda_{\mathrm{FS}}(\pmb{r}_{L}^{\prime}) \rangle .
    \label{eq:Pm_correlation_real}
\end{align}
We now perform the Fourier transform of  $\pmb{r}_{L}$ to $\pmb{K}_{L}$,
\begin{align}
   \delta \lambda_{\mathrm{FS}}  (\pmb{r}_L) =  \int \frac{ d^{3} \pmb{K}_L }{(2\pi)^{3} } \delta\lambda_{\mathrm{FS}} (\pmb{K}_{{L}})  
   e^{-i \pmb{K}_{L}\cdot \pmb{r}_{L}}.
\end{align}
The correlation function in Eq.~\eqref{eq:Pm_correlation_real} is expressed in terms of the power spectrum of $\langle  \delta \lambda_{\mathrm{FS}}(\pmb{K}_{L})   \delta \lambda_{\mathrm{FS}}(\pmb{K}_{L}^{\prime})\rangle \equiv  (2\pi)^{3} \delta^{(3)}(\pmb{K}_{L} + \pmb{K}_{L}^{\prime})P_{\delta \lambda_{\mathrm{FS}}}(\pmb{K}_{L} )$ as
\begin{align}
        \langle   \Delta   P_{\delta }(\pmb{k},\lambda_{\mathrm{FS}}(\pmb{r}_L);\pmb{r}_{L})   
        \Delta   P_{\delta}(\pmb{k},\lambda_{\mathrm{FS}}(\pmb{r}_L^{\prime});\pmb{r}_{L}^{\prime})\rangle 
         = 
         \left(\frac{d P_{\delta}}{d\lambda_{\mathrm{FS}} } \right)^2  \int \frac{d^{3}\pmb{K}_{L}}{(2\pi)^{3}}  
         P_{\delta \lambda_{\mathrm{FS}}}(\pmb{K}_L)e^{-i\pmb{K}_{L} \cdot (\pmb{r}_L - \pmb{r}_L^{\prime})}, 
\end{align}
which further leads to 
\begin{align}
     P_{PP}^{\delta}(\pmb{k}, \pmb{K}_{L}) =   
     \left(\frac{d P_{\delta}}{d\lambda_{\mathrm{FS}} } \right)^2   P_{\delta \lambda_{\mathrm{FS}}}(\pmb{K}_L)
\end{align}
with $\langle   \Delta   P_{\delta }(\pmb{k},\lambda_{\mathrm{FS}};\pmb{K}_{L})   \Delta   P_{\delta}(\pmb{k},\lambda_{\mathrm{FS}};\pmb{K}_{L}^{\prime})\rangle \equiv  (2\pi)^{3} \delta^{(3)}(\pmb{K}_{L} + \pmb{K}_{L}^{\prime})P_{PP}^{\delta}(\pmb{k}, \pmb{K}_{L}) $. 
Normalizing $\delta \lambda_{\mathrm{FS}}$ with respect to $\bar{\lambda}_{\mathrm{FS}}$ yields
\begin{align}
         P_{PP}^{\delta}(\pmb{k}, \pmb{K}_{L}) =  \lambda_{\mathrm{FS}}^2 \left(\frac{d {P}_{\delta}}{d\lambda_{\mathrm{FS}} } \right)^2   P_{\delta_\lambda}(\pmb{K}_L),
     \label{eq:modulating_matter_power_supp}
\end{align}
reproducing the long-range modulation of the matter power spectrum in Eq.~\eqref{eq:modulating_matter_power} by identifying $\delta $ with the matter density contrast.

\section{A concrete model }

In the main text we discussed the generation of kinetic isocurvature perturbations through a setup involving the decay of a heavy particle with a modulating decay rate. 
Here we elaborate on the consistency of the cosmological evolution in the setup.

We take a scalar $\phi$ decaying into a scalar dark matter $\chi$, and the modulating scalar field $\sigma$, in which the $\phi$ modes are all concentrated in the zero mode. 
The interaction term and the corresponding decay rate are 
\footnote{
$\phi$ can also decay into $\sigma$ and a pair of $\chi$. This decay mode is subdominant as long as $4\pi \sigma_i \gg m_\phi$.
}
\begin{align}
    \mathcal{L}_\mathrm{int} & = g \sigma \phi \chi^2 ~,
    \\ 
    \Gamma_{\phi} & \simeq \frac{1}{8\pi} \frac{g^2 \sigma_i^2}{m_{\phi}},
\end{align}
where $\sigma_i$ is the initial field value of $\sigma$. 
There are three temperature scales associated with the setup: The temperature $T_\mathrm{dec}$ when $\phi$ decays, the temperature $T_{\mathrm{osc}}$ when $\phi$ begins oscillations, and the temperature $T_{\mathrm{NR}}$ around which $\chi$ becomes nonrelativistic:
\begin{align}
    T_{\mathrm{dec}}  & \simeq \sqrt{\Gamma M_{P}}~,
    \\
    T_{\mathrm{osc}} & \simeq \left( \frac{90}{\pi^2 g_{*}} \right)^{1/4} \sqrt{m_{\phi} M_{P}} ~,
    \\ 
    T_{\mathrm{NR}} & \simeq T_{\mathrm{dec}} \frac{m_{\chi}}{m_{\phi}},
\end{align}
where $M_{P} \equiv (8\pi G)^{-1/2}$ and we assumed RD at $T_{\rm dec}$ and $T_{\rm osc}$, although the assumption is not crucial for the setup.
The dark matter density normalized by the entropy density, $\rho_{\chi} / s$, is
\begin{align}
    \frac{\rho_{\chi}}{s}  \simeq \frac{m_{\chi} n_{\chi}}{s} \simeq m_{\chi} \frac{n_{\phi}}{s} \sim \frac{m_{\chi} m_{\phi} \phi_{i}^2 }{g_{*} (m_{\phi} M_{P})^{3/2}},
\end{align}
where $\phi_{i}$ is the initial field value of $\phi$.
This value can be determined by identifying $\rho_{\chi} / s $ with the matter-radiation equality temperature $T_{\rm eq}$,  
\begin{align}
    \phi_{i} & \sim 
    \left(\frac{g_{*} T_\mathrm{dec} T_\mathrm{eq}}{T_\mathrm{NR}} \right)^{1/2} \frac{M_{P}^{3/4}}{m_{\phi}^{1/4}} 
    \nonumber\\
    & \sim 
    10^{15} \, \mathrm{GeV} \times \left( \frac{T_\mathrm{dec}}{10^{9 } \, \mathrm{GeV}} \right)^{1/2} 
    \left( \frac{10\, \mathrm{keV}}{T_\mathrm{NR}} \right)^{1/2 } \left( \frac{10^{8 } \, \mathrm{GeV}}{m_{\phi}} \right)^{1/4}. 
\end{align}
If the universe is matter dominated when $\phi$ begins oscillations, the required initial field value is larger.

In order for the modulation of the decay rate not to be suppressed, the modulating field $\sigma$ should start its oscillations only at $T<T_{\mathrm{dec}} $, which leads to the condition
\begin{align}
    m_{\sigma } < \frac{T_{\mathrm{dec}}^2}{M_P} = 1 \, \mathrm{GeV} \left(\frac{T_{\mathrm{dec}}}{10^9 \, \mathrm{GeV}} \right)^2.
    \label{smeq:sigma_mass_criteria}
\end{align}
The quantum correction to the quartic coupling $\phi^2 \sigma^2$ gives a mass to $\sigma$,
\begin{align}
    \delta m_{\sigma}^{\text{1-loop}} \sim \frac{g}{4\pi} \phi. 
\end{align}
Requiring this to satisfy Eq.~\eqref{smeq:sigma_mass_criteria} at $T_{\rm dec}$, we obtain an upper bound on the $g$ coupling: 
\begin{align}
    g\lesssim 10^{-8} \left(\frac{T_\mathrm{NR}}{10 \, \mathrm{keV}} \right)^{1/2} \left( \frac{m_{\phi}}{10^{8}\,\mathrm{GeV}} \right),
\end{align}
where we used $\phi \propto T^{3/2}$ after the beginning of the oscillations of $\phi$.

After the decay of $\phi$, $\sigma$ starts oscillations and is eventually dissipated at a temperature $T_{\rm dis}$.
The energy density of $\sigma$ at the time of dissipation should be sufficiently small, so that the cosmic perturbations produced by $\sigma$ through the curvaton mechanism~\cite{Linde:1996gt,Lyth:2001nq,Moroi:2001ct} should be smaller than the observed cosmic perturbations. 
The magnitude of the curvature perturbation produced by $\sigma$ is
\begin{align}
    \mathcal{R}_\sigma & \sim 
    \frac{\delta \sigma}{\sigma_i} \frac{\sigma_i^2}{M_P^2} \frac{\sqrt{m_\sigma M_P}}{T_{\rm dis}} 
    \nonumber\\
    & \sim 
    10^{-6} \times \frac{\delta \sigma}{\sigma_i} \frac{m_\phi}{10^8~{\rm GeV}} \left(\frac{T_{\rm dec}}{10^9~{\rm GeV}}\right)^2 \left(\frac{m_\sigma}{1~{\rm GeV}}\right)^{1/2} \frac{10^3~{\rm GeV}}{T_{\rm dis}} \left(\frac{10^{-8}}{g}\right)^2,
\end{align}
where we used the relation between $g \sigma_i$ and $T_{\rm dec}$. 
One can see that the curvature perturbations produced by $\sigma$ can be small enough for sufficiently large $T_{\rm  dis}$.
For example, if $\sigma$ has a coupling $A_H\sigma |H|^2$, where $H$ is the Standard Model Higgs, $\sigma$ is dissipated with a rate $\sim A_H^2/(8\pi T)$ and $T_{\rm dis}$ is given by
\begin{equation}
    T_{\rm dis}\sim 10^3~{\rm GeV} \left(\frac{A_H}{10^{-3}~{\rm GeV}}\right)^{2/3}.
\end{equation}
The quantum correction to the mass of $\sigma$ from $A_H$ can be small enough.

After $\sigma$ is thermalized, if its mass is below MeV, $\sigma$ will produce too large amount of $\Delta N_{\rm eff}$.
In fact, for $m_\sigma <$ MeV, if $\sigma$ continues to be in thermal equilibrium with the bath at $T\sim m_\sigma$, its entropy is transferred into the electron and photon, creating negative $\Delta N_{\rm eff}$.
If $\sigma$ is not in thermal equilibrium at $T\sim m_\sigma$, it will decay later while being nonrelativistic, creating even more $\Delta N_{\rm eff}$.
To avoid this, $m_{\sigma} \gg \mathrm{MeV}$ is required. This also limits $T_{\mathrm{dec}}$ to be above $ 10^{8} \, \mathrm{GeV}$. 




\end{document}